\documentclass[apj]{emulateapj}
\usepackage{apjfonts}
\usepackage{natbib}
\usepackage{graphicx}

\shorttitle{Discovery of VHE $\gamma$-ray emission from BL Lacertae}
\shortauthors{Albert et al.}

\begin{document}


 \title{Discovery of Very High Energy $\gamma$-ray emission from the Low-frequency peaked BL Lacertae object BL Lacertae}


\author{
 J.~Albert\altaffilmark{a}, 
 E.~Aliu\altaffilmark{b}, 
 H.~Anderhub\altaffilmark{c}, 
 P.~Antoranz\altaffilmark{d}, 
 A.~Armada\altaffilmark{b}, 
 C.~Baixeras\altaffilmark{e}, 
 J.~A.~Barrio\altaffilmark{d},
 H.~Bartko\altaffilmark{f}, 
 D.~Bastieri\altaffilmark{g}, 
 J.~K.~Becker\altaffilmark{h}
% ...
}
 \altaffiltext{a} {Universit\"at W\"urzburg, D-97074 W\"urzburg, Germany}
 \altaffiltext{b} {Institut de F\'\i sica d'Altes Energies, Edifici Cn., E-08193 Bellaterra, Spain}
 \altaffiltext{c} {ETH Zurich, CH-8093 Switzerland}
 \altaffiltext{d} {Universidad Complutense, E-28040 Madrid, Spain}
 \altaffiltext{e} {Universitat Aut\`onoma de Barcelona, E-08193 Bellaterra, Spain}
 \altaffiltext{f} {Max-Planck-Institut f\"ur Physik, D-80805 M\"unchen, Germany}
 \altaffiltext{g} {Universit\`a di Padova and INFN, I-35131 Padova, Italy}  
 \altaffiltext{h} {Universit\"at Dortmund, D-44227 Dortmund, Germany}
% ...


\begin{abstract}
The MAGIC collaboration observed BL Lacertae for 22.2 hr during 2005 August to December and for 26 hr during 2006 July to September. The source is the historical prototype and eponym of a class of low-frequency peaked  BL Lacertae (LBL) objects.
A very high energy (VHE) $\gamma$-ray signal was discovered with a 5.1 $\sigma$ excess in the 2005 data. Above 200 GeV, an integral flux of $(0.6\pm0.2)\times10^{-11}~{\rm cm}^{-2}~{\rm s}^{-1}$ was measured, corresponding to approximately 3\% of the Crab flux. The differential spectrum between 150 and 900 GeV is rather steep with a photon index of $-3.6\pm0.5$. 
The light curve shows no significant variability during the observations in 2005.
For the first time a clear detection of VHE $\gamma$-ray emission from an LBL object was obtained with a signal below previous upper limits. 
The 2006 data show no significant excess. This drop in flux follows the observed trend in optical activity.

\end{abstract}





\section{Introduction}

  BL Lacertae (1ES2200+420, $z$=0.069, \citealt{Miller78}) is the historical prototype of a class of powerful $\gamma$-ray emitters: ''BL Lac objects''. It belongs to an active galactic nucleus (AGN) subclass in which the jet is aligned very close to our line of sight. The mass of the supermassive black hole in the center is estimated to be $\sim 10^{8} M_{\sun}$~\citep{Wu04}.
Electromagnetic emission from this class of sources can be observed from the radio up to very high energy (VHE) $\gamma$-rays $(E>$100 GeV), with spectral energy distributions (SEDs) characterized by a two-bump structure. So far, most of the measured SEDs could be interpreted using scenarios of a simple leptonic origin. The lower energy bump is produced by synchrotron radiation of relativistic electrons, while the higher energy bump originates from inverse Compton (IC) scattering of the same electron population accelerated in the jet. The target photons for IC scattering could be either synchrotron photons (synchrotron self-Compton scattering, SSC) or external photons either from the broad-line emission region or from the accretion disc (external inverse Compton scattering, EC). If both photon populations contribute to the IC emission, the corresponding IC component may have a double local-peak structure~\citep[e.g.][]{Ghi98}. When the synchrotron emission peak is located in the sub-millimeter to optical band, the objects are classified as ''Low-frequency peaked BL Lacs" (LBLs), whereas in ''High-frequency peaked BL Lacs'' (HBLs) the synchrotron peak is located at UV to X-ray energies~\citep{Pad95, Fos98}. BL Lac objects often show a strong flux variability down to timescales of a few minutes for the highest energy range~\citep{Gai96, MAGIC07a}. It should be noted that other models, based e.g.\,on the acceleration of hadrons~\citep{Man93, Muc01}, could also explain the SEDs of BL Lac objects.

Up to now, there are 13 well-established extra-galactic VHE $\gamma$-ray sources. Twelve are classified as HBL and one is the Fanaroff-Riley class I radio galaxy, M87. BL Lacertae is classified as an LBL object with a synchrotron peak frequency of $2.2 \times 10^{14}$ Hz~\citep{Sam99}, and is one of the best-studied objects in the various energy bands.
In 1998 and 2001, VHE $\gamma$-ray detection of the LBLs 3C66A and BL Lacertae were claimed~\citep{Nes98, Nes01}. Subsequent observations with more sensitive instruments have not, however, revealed any evidence for VHE $\gamma$-ray emission consistent with these claims.
No VHE $\gamma$-ray emission from any LBLs has previously been confirmed.

\citet{Vil04} have presented long-term light curves in optical and radio emission over 30 years and reported cross-correlation between the optical light curve and radio hardness ratio with some delay in radio emission. They have claimed evidence of an $\sim$8 yr periodicity in radio but less evidence in optical. Several authors~\citep[e.g.][]{Hag97, Sti03} have also reported periodic and quasi-periodic variations in the optical and radio light curves.

   \begin{figure}
   \centering
      \includegraphics[width=6.7cm]{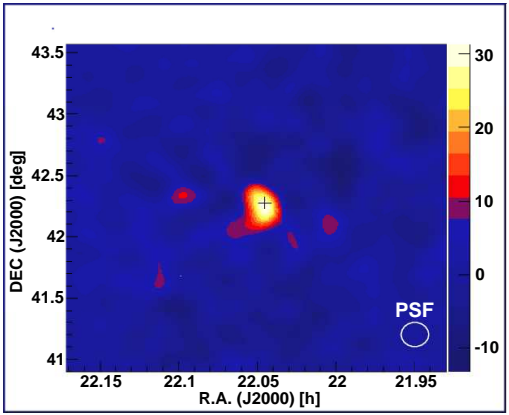}
                      \caption{Sky map of the region around the position of BL Lacertae (black cross) for reconstructed $\gamma$ events with $>$ 350 photo-electrons (corresponding to an energy threshold of about 200 GeV) from the 2005 observations.}	
         \label{sky}
   \end{figure}
%
%

Gamma-ray observations by EGRET resulted only in an upper limit of 1.4 $\times$ 10$^{-7}$\ ${\rm cm}^{-2}\ {\rm s}^{-1}$ until 1995. Later, EGRET observed $\gamma$-rays above 100 MeV at a flux level of (4.0$\pm 1.2) \times 10^{-7}\ {\rm cm}^{-2}\ {\rm s}^{-1}$ with 4.4 $\sigma$ significance~\citep{Cat97}.
During an optical outburst in 1997, a $\gamma$-ray flare was measured with 10 $\sigma$ significance by EGRET at a flux level of (1.72$\pm$0.42) $\times$ 10$^{-6}$ ${\rm cm}^{-2}\ {\rm s}^{-1}$, 12 times higher than the previous upper limit~\citep{Blo97}. The EC, in which the broad-line emission flux has been Comptonized, has been suggested for the interpretation of the $\gamma$-ray emission in the 1997 flare~\citep[e.g.][]{Mad99, Rav02}.

In the VHE $\gamma$-ray range, the Crimean Observatory has claimed a detection with 7.2 $\sigma$ significance~\citep{Nes01},
while HEGRA, observing in the same period, obtained only a much lower upper limit~\citep[][see details in Discussion]{Kra03}. Other past observations of this target resulted in upper limits only~\citep{Cat97, Kra03, Aha04, Hor04}.

In this Letter we report about the discovery of VHE $\gamma$-ray emission from BL Lacertae in 2005. Simultaneous observation in the optical band in 2005 and 2006 permitted a search for correlations between the optical and VHE $\gamma$-ray activities.

\section{Observations and Data Analysis}


BL Lacertae (R.A.$22^{\rm h}02^{\rm m}43.3^{\rm s}$,\,decl.$+42^{\circ}16'40''$;\,J2000.0) was observed with the 17m diameter MAGIC (Major Atmospheric Gamma Imaging Cerenkov) telescope on the Canary Island of La Palma (N28.2$^{\circ}$, W17.8$^{\circ}.8$, 2225 m a.s.l.). The MAGIC telescope is the world's largest Imaging Atmospheric Cerenkov Telescope (IACT).
The telescope parameters and performance are described in \citet{Baix04} and \citet{Cor05}.
The source was observed for 22.2 hr from 2005 August until December. The telescope was pointing directly onto the object, recording so-called ON data. The background was estimated from observations of regions where no $\gamma$-rays are expected, which we define as OFF data, which were taken with sky conditions similar to those of ON data.

The raw data were first calibrated~\citep{Gau05}, and then processed with the standard MAGIC analysis and reconstruction software~\citep{Bre03}.
Data runs with anomalous trigger rates due to bad observation conditions were rejected from further analysis. The remaining ON data corresponded to 17.8 hr, while the OFF data corresponded to 57.2 hr. Image parameters of the raw data~\citep{Hillas} were calculated and compared for the ON and OFF data in order to check their consistency; excellent agreement was found. The selected data samples before the $\gamma$/hadron separation are completely dominated by hadron events and the expected admixture of $\gamma$-ray events are below the statistical fluctuations of the data.

   \begin{figure}
   \centering
\includegraphics[width=8.cm]{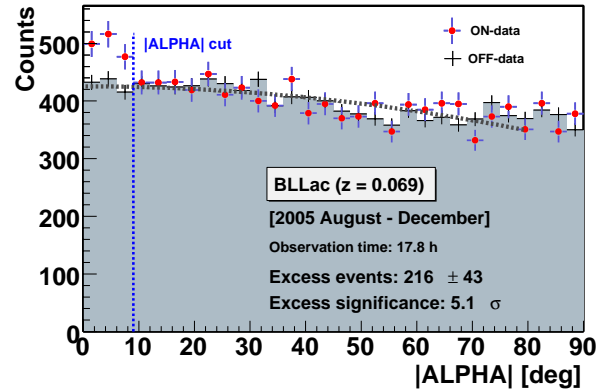}
   \caption{ALPHA-distribution of the 2005 data. The filled circles represent ON data. The light crosses correspond to normalized OFF data and a dotted curve describes a second order polynomial fit to the distribution of the OFF data. The vertical line indicates the ALPHA selection condition, which yields a total excess of 216 events at a significance level of 5.1 $\sigma$.
   }
   \label{alpha}
   \end{figure}

%

In order to reject most hadronic background while still conserving the majority of the $\gamma$-candidates, a multi-tree classifier algorithm based on the "Random Forest (RF)" method~\citep{Bre01,Bock04} was used for the $\gamma$/hadron separation. The selection conditions were trained with Monte Carlo simulated $\gamma$-ray samples~\citep{Kna04, Maj05} and a sample of experimental background events.
In the RF method for each event the so-called HADRONNESS parameter (H) was calculated from a combination of all the image parameters except the ALPHA parameter. ALPHA is the angle between the shower image principal axis and the line connecting the image center of gravity with the camera center. H assigns to each event a number between 0 and 1 of being more hadron-like (high H) or $\gamma$-ray like (low H values). The selection of events with low H value enriches $\gamma$-ray events in the surviving data sample. Finally, the image parameter ALPHA was calculated and the ALPHA distribution for the ON and OFF data compared after normalizing them on the area between $20^{\circ}$ and $80^{\circ}$. Any $\gamma$-ray signal should show up as an excess at small ALPHA values. The number of background events in the ON distribution at small ALPHA was determined by a second order polynomial fit (without linear term) to the ALPHA distribution of the normalized OFF data. For our analysis we used an ALPHA cut of $9^{\circ}$~\citep[see also][]{MAGIC06a, MAGIC07b, MAGIC06b}.

The cuts in H and ALPHA were optimized using data samples from Crab Nebula observations at comparable zenith angles. The significance of an excess visible in the ALPHA distribution was calculated according to Eq.17 in~\citet{LiMa}. The energy of the $\gamma$-ray events was also reconstructed by means of the RF method with MC $\gamma$-samples. The average energy resolution is estimated to be about 24\% rms.

Optical $R$-band observations were provided by the Tuorla Observatory Blazar Monitoring Program\footnote{more information at http://users.utu.fi/kani/1m/} with the 1.03 m telescope at the Tuorla Observatory, Finland, and the 35 cm KVA telescope on La Palma, Canary Islands. The magnitudes were then converted to linear fluxes using formula F[Jy] $=3080.0\times10^{-({\rm mag}/2.5)}$.

Radio observations were also performed with UMRAO\footnote{UMRAO is partially supported by a series of grants from the NSF and by funds from the University of Michigan.} at 4.8, 8.0 and 14.5 GHz, and at 37 GHz with the Mets\"ahovi Radio Observatory.

   \begin{figure}
   \centering
   \includegraphics[width=7.5cm]{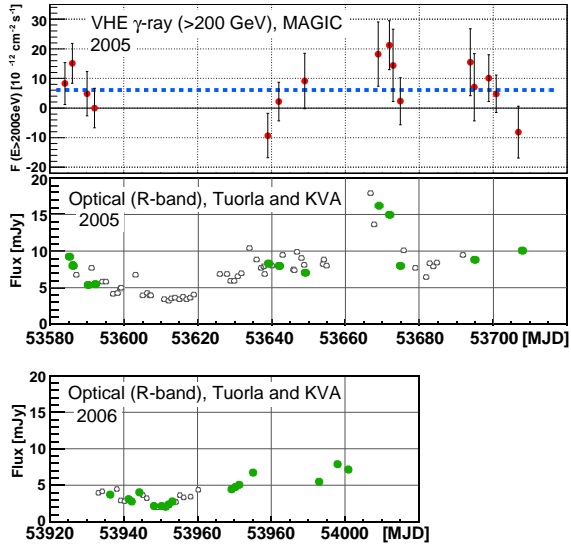}
                    \caption{Light curve of VHE $\gamma$-ray ($>$ 200 GeV) flux as measured with the MAGIC telescope during 2005 [top]. A dotted horizontal line represents the average flux. Light curve of optical ($R$-band) flux during 2005 [middle] and 2006 [bottom] as measured with the 1.03m Tuorla and the 35 cm KVA telescope. In the optical light curve, the filled points represent the optical flux when simultaneous MAGIC observations were carried out. The measured average flux of those points is 9.2 mJy for 2005 and 4.2 mJy for 2006
              }
         \label{LC}
   \end{figure}


\section{Results}

 Figure~\ref{sky} shows the local sky map for the 2005 $\gamma$-ray candidates using the DISP method~\citep{Les01}. Only events $>$ 350 photo-electrons were used, corresponding to an energy threshold of about 200 GeV $\gamma$-ray energy. The map was produced from the excess events distribution smoothed with a 2-D Gaussian of $0.1^{\circ}$.
The black cross marks the nominal position of BL Lacertae. The small offset and the extension of the image are comparable to the telescope point spread function (PSF, $0.1\degr$) and the telescope pointing error ($1.5\arcmin$).

In Figure~\ref{alpha}, the ALPHA distribution is shown. 
An excess of 216 events over 1275.6 normalized background events yields a significance of 5.1 $\sigma$ for data above 350 photo-electrons. 
Two mostly independent analyses confirmed the result.

The 2005 VHE $\gamma$-ray and the 2005 and 2006 optical light curves are shown in Figure~\ref{LC}. 
No significant evidence of flux variability in VHE $\gamma$-rays was found in the 2005 data. 
The derived average integral flux is $F(E>200\ {\rm GeV})= (0.6 \pm 0.2) \times 10^{-11}\ {\rm cm}^{-2}\ {\rm s}^{-1}$ $ (\chi^{2}/\rm{dof} = 16.3/15)$, which corresponds to about 3\% of the Crab Nebula flux as measured by the MAGIC telescope~\citep{Wag05}.
In the optical light curve, the contribution from the host galaxy (1.38 mJy) was subtracted. The optical light curve shows a flare around the end of 2005  October (MJD $\sim$53,670). Also radio light curves at 37 GHz (A. L\"ahteem\"aki 2006, private communication) and 14.5 GHz (UMRAO) suggest some minor flaring activity starting in 2005 November.

\begin{figure}
   \centering
  \includegraphics[width=7.15cm]{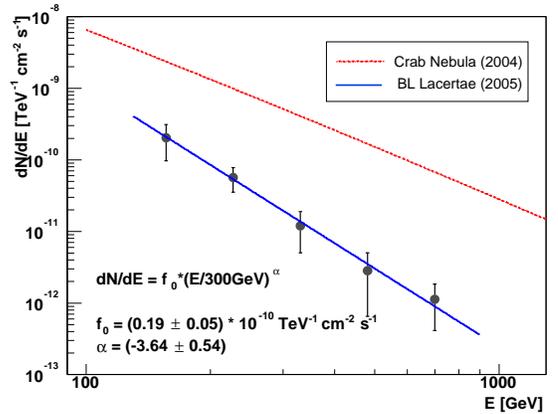}
                       \caption{Differential energy spectrum of the 2005 BL Lacertae data. The solid line represents a power-law fit to the measured spectrum. The fit parameters are listed in the figure. For comparison, the line fitted to the measured MAGIC CRAB spectrum using a power law with a changing photon index is shown by a dashed line~\citep{Wag05}.}
         \label{sp}
   \end{figure}

The reconstructed differential energy spectrum (Fig.~\ref{sp}) is well described by a simple power law:
\begin{displaymath}
 \frac{dN}{dE} = (1.9 \pm 0.5) \times 10^{-11} \left( \frac{E}{300\ {\rm GeV}}\right)^{-3.6\pm0.5} \frac{{\gamma}}{{\rm TeV}\ {\rm s}\ {\rm cm}^2}
\end{displaymath}
Because of the relatively steep slope, the systematic errors are estimated to be $\sim$ 50\% for the absolute flux level and 0.2 for the spectral index. 

Follow-up observations were carried out from 2006 July to September for 26.0 hr, using the so-called wobble mode~\citep{Dau97}, where the object was observed at an $0.4^{\circ}$ offset from the camera center; 25.0 hr of the data passed all quality selection for the analysis. The DISP method was applied for the reconstruction of the shower direction with the final cut on the $\theta^{2}$ parameter (the squared angular distance between the nominal source position and the reconstructed $\gamma$-ray direction). However, no significant excess could be found in the 2006 data.


%
%

\section{Discussion}

The Whipple 10 m telescope observed BL Lacertae for 39.1 hr in 1995 and derived a flux upper limit above 350 GeV at 3.8\% of Crab~\citep{Hor04}. HEGRA derived an upper limit above 1.1 TeV at 28\% of Crab with 26.7 hr observation~\citep{Aha04}. These upper limits are consistent with our results. On the other hand, \citet{Nes01} claim VHE $\gamma$-ray detection using the GT 48 telescopes of the Crimean Astrophysical Observatory from data taken in summer 1998. The reported integral $\gamma$-ray flux is $F(E>1\ {\rm TeV}) = (2.1 \pm 0.4) \times 10^{-11}\ {\rm cm}^{-2}\ {\rm s}^{-1}$, which is two orders of magnitude higher than the extrapolated value from this study. During the same period, 1998 July to August, no significant signal was found by HEGRA, and their reported flux upper limit in the same energy band is 7 times lower than the Crimean result~\citep{Kra03}. The VHE $\gamma$-ray detection reported by Neshpor et al. cannot be explained without a remarkably huge and a very rapid flux variation offset by a few hours in consecutive nights from the HEGRA observation. In the case of a leptonic origin of the $\gamma$-ray emission, such a flare would possibly coincide with high activity in the optical as in the outburst of 1997 July, when the increase in flux was observed both in the optical and X-ray to $\gamma$-ray bands~\citep{Blo97, Tan00}. However, no increased optical activity was detected during the Crimean observation period ($m_p = 13.5-14.6$, while $13.0-14.6$ in 2005). Throughout the EGRET observations for BL Lacertae, such a notably huge and rapid flare feature was never reported in the high energy $\gamma$-ray emission, which is considered to have the same origin as the VHE $\gamma$-ray emission regardless of the scenarios for the emission origin.

 In Figure~\ref{LC}, filled circles in the optical light curve represent simultaneous observations with the MAGIC telescope, accepting a $\pm$1 day offset with respect to the MAGIC observations. 
In 2005, there are 12 out of 16 nights with simultaneous observations (average flux: 9.2 mJy), while 16 out of 23 nights in 2006 (average flux: 4.2 mJy) have coinciding observations. 
The absence of a significant excess of VHE $\gamma$-rays in the 2006 data indicates that the VHE $\gamma$-ray flux in 2006 was significantly lower than the flux level in 2005. 
In summary our results show similar tendencies both in the optical and the VHE $\gamma$-ray flux variations. Similarly, the $\gamma$-ray activity seen by EGRET observations in 1997 showed a strong correlation with optical activity. Such a correlation is expected from leptonic origin scenarios~\citep{Blo97}.
Details of the 2006 data and radio data will be discussed elsewhere.

   \begin{figure}[t]
   \centering
    \includegraphics[width=8.2cm]{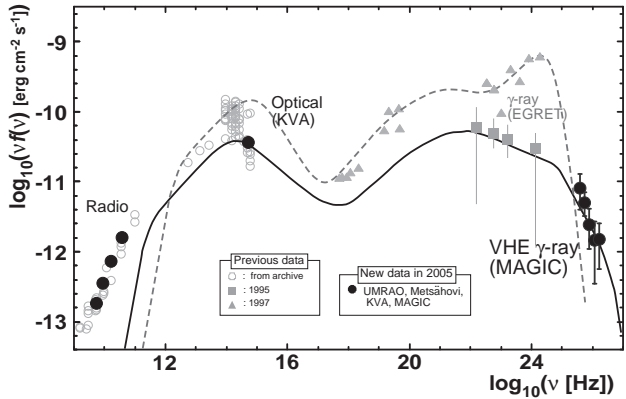}
     \caption{SED of BL Lacertae. Black filled circles represent simultaneous 2005 data of KVA and MAGIC as well as radio data taken by UMRAO and Mets\"ahovi. 
Gray color points describe old measurements (see detail in the inlay). 
The lines are taken from \citet{Rav02}. The solid line represents a one-zone SSC model for the 1995 data, the dotted line is produced with SSC and EC components for the 1997 flare data.}
         \label{SED}
   \end{figure}
%

 Figure~\ref{SED} shows the SED of BL Lacertae with results of this work and some previous data and model calculations by \citet{Rav02}. The VHE $\gamma$-ray points are corrected for the extra-galactic background light (EBL) absorption using the "Low" EBL model of \citet{Kne04}. Our optical and VHE $\gamma$-ray points agree well with the solid line, which was derived using a one-zone SSC model on the 1995 data, whereas some deviations can be seen from the dotted line, which describes the 1997 flare data and involves SSC as well as EC components~\citep{Rav02}. 
To describe our result such an additional EC component is not necessarily required.

BL Lacertae is the first LBL object with a clear detection of VHE $\gamma$-ray emission. The results of this work indicate that VHE $\gamma$-ray observations during times of higher optical states can be more efficient.
Future long term monitoring of VHE $\gamma$-ray emission could provide detailed information of a possible periodicity predicted by e.g.~\citet{Sti03} and correlations with other wavelengths.
Due to the observed steep spectrum, lowering the energy threshold of IACTs (e.g.\,the upcoming MAGIC-II project) would significantly increase the detection prospects for this new source class.

\acknowledgements
We would like to thank the IAC for the excellent working conditions at the ORM. The support of the German BMBF and MPG, the Italian INFN, the Spanish CICYT, the Swiss ETH and the Polish MNil is gratefully acknowledged.
We thank H. Aller and M. Aller (UMRAO), and A. L\"ahteenm\"aki (Mets\"ahovi) for providing the radio data.




\clearpage

\end{document}